\documentclass[a4paper,12pt]{article}

\usepackage{bbold}
\usepackage{float}
\usepackage{cancel}
\usepackage{subfig}
\usepackage{amsmath, amssymb, setspace,cite}
\usepackage{graphicx}
\usepackage{hyperref}
\usepackage{color}
\usepackage{soul}
\usepackage{xcolor}
\usepackage{multirow}

\vspace{0.6cm}
\date{\today} 

\usepackage{tabularx}
\usepackage{longtable} 
\usepackage{multicol}
\usepackage{multirow}

\renewcommand{\arraystretch}{1.2}

\definecolor{dgreen}{rgb}{0.0, 0.5, 0.0}

\usepackage{marginnote}

\begin{document}

\hfill {\tt CERN-TH-2023-208}

\def\thefootnote{\fnsymbol{footnote}}

\begin{center}
\Large\bf\boldmath
\vspace*{1.cm} 
Beyond the Standard Model prospects for kaon physics\\ at future experiments\unboldmath
\end{center}
\vspace{0.6cm}

\begin{center}
G.~D’Ambrosio$^{1}$\footnote{Electronic address: gdambros@na.infn.it}, 
F.~Mahmoudi$^{2,3,4,}$\footnote{Electronic address: nazila@cern.ch}, 
S. Neshatpour$^{2,}$\footnote{Electronic address: s.neshatpour@ip2i.in2p3.fr}\\
\vspace{0.6cm}
{\sl $^1$INFN-Sezione di Napoli, Complesso Universitario di Monte S. Angelo,\\ Via Cintia Edificio 6, 80126 Napoli, Italy}\\[0.4cm]
{\sl $^2$Universit\'e de Lyon, Universit\'e Claude Bernard Lyon 1, CNRS/IN2P3, \\
Institut de Physique des 2 Infinis de Lyon, UMR 5822, F-69622, Villeurbanne, France}\\[0.4cm]
{\sl $^3$Theoretical Physics Department, CERN, CH-1211 Geneva 23, Switzerland}\\[0.4cm]
{\sl $^4$Institut universitaire de France (IUF)}
\end{center}

\renewcommand{\thefootnote}{\arabic{footnote}}
\setcounter{footnote}{0}

\vspace{1.cm}
\begin{abstract}
Rare kaon decays offer a powerful tool for investigating new physics in $s\to d$ transitions. Currently, many of the interesting decay modes are either measured with rather large uncertainties compared to their theoretical predictions or have not yet been observed. The future HIKE programme at CERN will provide unprecedented sensitivity to rare kaon decays, allowing for strong constraints on new physics scenarios with lepton flavour universality violation. We present the overall picture that emerges from a study of the different decay modes with a global analysis considering projections based on the HIKE programme, both with and without 
KOTO-II future measurements. We also highlight the most relevant decays and identify that in addition to the ``golden channel'',  $K^+\to\pi^+\nu\bar\nu$, the rare $K_L\to\pi^0 \ell\bar\ell$ decay, especially in the electron sector offers strong constraints on short-distance physics. 
\end{abstract}
\newpage

\section{Introduction}

In the ever-evolving landscape of particle physics, the study of kaons occupies certainly an important role, offering new insights and an improved sensitivity to physics beyond the Standard Model (SM). At present, the field of kaon physics finds itself at a crossroads, characterised by ongoing experiments, such as NA62 and KOTO, at the highest level of activity studying rare kaon decays. The future trajectory of kaon physics unfolds with the contemplation of upcoming experiments such as HIKE~\cite{HIKE:2022qra,HIKE-Proposal}, KOTO-II~\cite{Aoki:2021cqa}, and LHCb upgrade II~\cite{LHCb:2018roe,AlvesJunior:2018ldo}, which can not only advance the current experiments but also anticipate new avenues for exploration. 

In this context, it is well-founded to address the achievements of ongoing experiments in a global analysis to assess the current sensitivity to new physics parameters, together with discussing in detail the role of each of the kaon decays that have been measured, and to provide a detailed analysis of future sensitivities of the upcoming experiments~\cite{Ahdida:2023okr}. This is precisely the aim of this paper.

A specific category within the realm of rare processes, known as ``golden modes'', holds a distinct role in the indirect exploration of new physics. Golden modes involve rare decays where we can measure theoretically clean observables. This allows us to uncover contributions from new physics that compete favourably with SM processes. Within the framework of the future experiment programme, the focus is on precisely measuring two key golden modes: $K^+\to \pi^+ \nu \bar{\nu}$ and $K_L \to \pi^0 \nu \bar{\nu}$. 
The accurate measurement of the branching ratios for these decay processes serves as a model-independent benchmark, providing reliable constraints for beyond the Standard Model (BSM) scenarios.

The decay $K^+\to\pi^+\nu\nu$ is very precisely measured but it does not distinguish between flavour families. However, its interplay with other rare kaon decays provides strong constraints for lepton flavour universality violating (LFUV) effects. Other rare kaon decays include $K_L\to\pi^0 \ell\bar{\ell}$ which is planned to be measured within the HIKE programme for both electron and muon channels. This decay mode although not theoretically as clean as the golden modes, can offer a strong probe of new physics.
In addition, the $K_L\to\mu\bar{\mu}$ decay has been measured with better than 2\% uncertainty, and although the long-distance dominated SM prediction has rather large uncertainty and depends on the sign of the two-photon contribution, it still offers valuable information on short-distance physics. Unlike the $K_L\to\mu\bar\mu$ decay, which exhibits sign ambiguity, the $K_S\to\mu\bar\mu$ decay does not. However, its experimental upper bound is approximately two orders of magnitude larger than the SM prediction, a situation that the upcoming LHCb upgrade aims to improve~\cite{LHCb:2018roe}. 
Another relevant decay mode is $K^+\to\pi^+ \ell\bar\ell$ which although currently lacks a reliable theoretical prediction can test lepton flavour universality violation. 

Currently, the primary obstacle in investigating BSM models, and more broadly, in the quest for new physics using kaons, lies in the limited statistical precision of kaon decay measurements. To a lesser extent, this limitation is also influenced by theoretical uncertainties in the SM predictions. Encouragingly, efforts are underway to reduce these uncertainties in the coming years, and we address in this study their impacts in global fits to new physics parameters. 

We examine the potential for rare kaon measurements to reveal departures from the accidental symmetries of the SM (LFUV), in upcoming experiments. We evaluate different benchmark points, incorporating the final precision expected from NA62, followed by the target precision for HIKE Phase 2, and finally, we include the ultimate precision anticipated for KOTO-II within the same timeframe as the HIKE programme.

\noindent The results presented in this work are obtained using the SuperIso public programme~\cite{Mahmoudi:2007vz,Mahmoudi:2008tp,Mahmoudi:2009zz,Neshatpour:2021nbn}.

The paper is organised as follows: In Section~\ref{sec:2} we present briefly the theoretical framework and the relevant observables that we consider in this study. Section~\ref{sec:global} presents the global analyses and our main results for future experiment prospects. Finally, our conclusions are given in Section~\ref{sec:conc}.

\section{Theoretical framework}
\label{sec:2}

The $s \to d$ transitions are parameterised using the following effective Hamiltonian:

\begin{equation}
\label{eq:Heff}
\mathcal{H}_{\rm eff}=-\frac{4G_F}{\sqrt{2}}\lambda_t^{sd}\frac{\alpha_e}{4\pi}\sum_k C_k^{\ell}O_k^{\ell}\,.
\end{equation}
Here $\lambda_t^{sd}\equiv V^*_{ts}V_{td}$, and the relevant operators are
\begin{eqnarray}
{O}_9^{\ell} &=& (\bar{s} \gamma_\mu P_L d)\,(\bar{\ell}\gamma^\mu \ell)\,,\nonumber\\ \label{eq:operators}
{O}_{10}^{\ell} &=& (\bar{s} \gamma_\mu P_L d)\,(\bar{\ell}\gamma^\mu\gamma_5 \ell)\,,\\
{O}_L^{\ell} &=& (\bar{s} \gamma_\mu P_L d)\,(\bar{\nu}_\ell\,\gamma^\mu(1-\gamma_5)\, \nu_\ell)\,,\nonumber
\end{eqnarray}
with $P_L=(1-\gamma_5)/2$. 
The Wilson coefficients $C_k^{\ell}$ are parameterised as:
\begin{equation}
C_k^{\ell} = C_{k,{\rm SM}}^{\ell}+ \delta C_{k}^{\ell}\,.
\end{equation} 
In general, besides the operators given in Eq.~(\ref{eq:operators}), new physics effects also contribute via (pseudo)scalar operators and those with right-handed quark currents which are not relevant for the SM. For this work, we only consider the above-mentioned subset, and consider new physics contributions in the chiral basis and assume the neutral leptons to be connected to their charged counterparts via SU(2)$_{\rm L}$ gauge symmetry, such that $\delta C_L^\ell \equiv \delta C_9^\ell= -\delta C_{10}^\ell$.
Furthermore, we investigate lepton flavour universality violation, considering new physics effects for electrons to be different compared to muons and taus,  $\delta C_L^e \neq \delta C_L^\mu (= \delta C_L^\tau)$.

\subsection{Observables considered in the global analysis}
\label{sec:2.1}
In this work, we consider rare kaon decays which are sensitive to new physics effects and for which a reliable theoretical prediction exists in the SM. The seven observables we consider are (see Refs.~\cite{DAmbrosio:2022kvb,Neshatpour:2022fak} for more detailed descriptions):

\begin{itemize}
\item The branching ratio of $K^+\to \pi^+ \nu\bar\nu$ and $K_L\to \pi^0 \nu\bar\nu$ which are theoretically very clean~\cite{Bobeth:2017ecx,Buchalla:1995vs,Buchalla:1996fp,Buchalla:1998ba,Isidori:2005xm,Mescia:2007kn,Misiak:1999yg,Brod:2010hi} and predicted with less than 8 and 12\% uncertainties, respectively. 
The dominant short-distance (SD) contributions in these decays make them very sensitive probes of new physics effects. However, experimentally the branching ratios are measured as the sum over the three neutrino flavours and it is not possible to differentiate between new physics contributions to electrons, muons and taus. On the experimental side, $K^+\to\pi^+\nu\bar{\nu}$ has already been measured by NA62 with $\sim40\%$ precision which will be improved to 20\% by the end of its run in 2025. The HIKE programme plans to measure this decay with an impressive 5\% precision at phase 1. On the other hand, $K_L\to\pi^0\nu\bar{\nu}$ has not yet been observed and the current upper bound by KOTO is two orders of magnitude larger than its SM prediction.

\item Lepton flavour universality violating effects in the $K^+ \to \pi^+ \ell \bar \ell$ decays. For this channel, although a precise theoretical determination is not yet  available~\cite{DAmbrosio:1998gur,Gilman:1979ud, Ecker:1987hd, Ecker:1987qi, DAmbrosio:1994fgc, Ananthanarayan:2012hu}, the comparison of the form factors of the electron and the muon modes offers a test on LFUV effects~\cite{Crivellin:2016vjc}.
For this decay mode any disagreement among the experimental determination of the form factor parameters $a_+^{\mu\mu}$ and $a_+^{ee}$ indicates short-distance LFUV effects among electrons and muons via $a_+^{\mu\mu}-a_+^{ee} = - \sqrt{2}\,{\rm Re}\left[V_{td}V^*_{ts} (C^{\mu}_{9}-C^{e}_{9}) \right] $.
Experimentally, the $a_+^{ee}$ has been measured by E865 and NA48/2 with the combination given in~\cite{DAmbrosio:2018ytt} and $a_+^{\mu\mu}$ was recently measured by NA62. The HIKE programme aims to reduce the uncertainty in $a_+^{ee} - a_+^{\mu\mu}$ to less than half of what is currently measured.

\item The branching ratios of $K_L \to \mu\bar\mu$ and $K_S\to\mu\bar\mu$. On the theory side, in these decays, the large long-distance contributions are dominant resulting in large theoretical uncertainties~\cite{DAmbrosio:1986zin,Ecker:1991ru,GomezDumm:1998gw,Knecht:1999gb, Isidori:2003ts, DAmbrosio:2017klp, Mescia:2006jd,Quigg:1968zz,Martin:1970ai,Savage:1992ac,DAmbrosio:1992zqm,DAmbrosio:1996lam,Valencia:1997xe,DAmbrosio:1997eof,Cirigliano:2011ny,Colangelo:2016ruc}. However, short-distance contributions can lead to sizeable effects, making it possible to extract constraints on new physics parameters. 
This is especially relevant for $K_L\to\mu\bar\mu$ which has been measured with less than 2\% uncertainty. The predicted BR($K_L\to\mu\bar\mu$) depends on whether the short-distance and long-distance contributions interfere constructively or destructively which depends on the unknown sign of 
${\cal A}(K_L\to\gamma\gamma)$.\footnote{In this work, for the $K_L\to\mu\bar\mu$ decay, unless otherwise stated, for the long-distance contribution via ${\cal A}(K_L\to\gamma\gamma)$, we consider the sign to be both positive and negative, denoted by LD$+$ and LD$-$, respectively.} 
For $K_S\to\mu\bar\mu$, the situation is different, while the theory prediction is not dependent on the sign ambiguity, on the experimental side, currently the upper bound by LHCb is about two orders of magnitude larger than the SM prediction making it difficult to obtain constraints on short-distance physics.    

\item The branching ratio of $K_L \to \pi^0 \ell \bar\ell$ both in the electron and the muon channels.
This decay mode has the following main contributions~\cite{Buchalla:2003sj,Isidori:2004rb,Mescia:2006jd,Martin:1970ai, Ecker:1987qi, Donoghue:1987awa, Ecker:1987hd, Ecker:1987fm, Sehgal:1988ej, Flynn:1988gy, Cappiello:1988yg, Morozumi:1988vy, Ecker:1990in, Savage:1992ac, Cappiello:1992kk, Heiliger:1992uh, Cohen:1993ta, Buras:1994qa, DAmbrosio:1996kjn, Donoghue:1997rr, KTeV:1999gik, Murakami:1999wi, KTeV:2000amh, Diwan:2001sg, Gabbiani:2001zn, Gabbiani:2002bk, NA48:2002xke}: \emph{1)} The CP-conserving two-photon contribution via $K_L\to\pi^0\gamma^*\gamma^*\to\pi^0\ell\bar\ell$. \emph{2)} The direct CP-violating term sensitive to short-distance physics and proportional to $\lambda_t$. \emph{3)} The indirect CP-violating contributions via $K_S \to \pi^0\ell\bar\ell$ which are proportional to the CP violating parameter $\epsilon$. \emph{4)} The interference between the latter two contributions which can be both destructive and constructive.\footnote{All through this paper, for the  $K_L\to\pi^0\ell\bar\ell$ decay, we assume constructive interference which is theoretically more favoured than destructive interference.}
The electron mode is more sensitive to new physics effects compared to its muon counterpart which is due to the larger available phase-space as well as the vanishing two-photon contribution for the electron mode. 
On the experimental side, current bounds are one order of magnitude larger than the SM predictions. 

\end{itemize}

The prediction and experimental measurements or upper bounds are collected in the second and third columns of Table~\ref{tab:data}.

\vspace*{1.cm}
\begin{table}[!t]
\vspace{-5mm}
\renewcommand{\arraystretch}{1.39}
\begin{center}
\setlength\extrarowheight{1pt}
\scalebox{0.65}{
\hspace*{-2mm}
\begin{tabular}{llll|ccc}\hline\hline
\bf{Observable} & \bf{SM prediction}& \bf{Experimental results} & \bf{Reference}& \bf{NA62 final} & \textbf{HIKE Phase 2} & \textbf{HP2 + KOTO-II} \\ \hline
BR$(K^+\to \pi^+\nu\bar\nu)$    & $(7.86 \pm 0.61)\times 10^{-11}$  & $(10.6^{+4.0}_{-3.5} \pm 0.9 ) \times 10^{-11}$ & \cite{NA62:2021zjw}& 20\%  & 5\% & 5\% \\
BR$(K^0_L\to \pi^0\nu\nu)$  & $(2.68 \pm 0.30) \times 10^{-11}$ & $ <3.0\times 10^{-9} $ @$90\%$ CL & \cite{Ahn:2018mvc}& current & current & $20\%$\\
LFUV($a_+^{\mu\mu}-a_+^{ee}$)&\multicolumn{1}{c}{0}&$-0.014\pm 0.016$&\cite{E865:1999ker,NA482:2009pfe,DAmbrosio:2018ytt,NA62:2022qes}& current &$\pm0.007$ &$\pm0.007$\\
BR$(K_L\to \mu\mu)$ ($+$)   & $(6.82^{+0.77}_{-0.29})\times 10^{-9}$    & \multirow{2}{*}{$(6.84\pm0.11)\times 10^{-9}$} & \multirow{2}{*}{\cite{ParticleDataGroup:2020ssz}} &  \multirow{2}{*}{current} & \multirow{2}{*}{1\%} & \multirow{2}{*}{1\%} \\
BR$(K_L\to \mu\mu)$ ($-$)   &  $ (8.04^{+1.47}_{-0.98})\times 10^{-9}$    &  & & 
\\
BR$(K_S\to \mu\mu)$         & $(5.15\pm1.50)\times 10^{-12}$    & $ < 2.1(2.4)\times 10^{-10}$ @$90(95)\%$ CL & \cite{LHCb:2020ycd} & current & current & current \\
BR$(K_L\to \pi^0 ee)(+)$         & $(3.46^{+0.92}_{-0.80})\times 10^{-11}$    & \multirow{2}{*}{$ < 28\times 10^{-11}$ @$90\%$ CL} & \multirow{2}{*}{\cite{KTeV:2003sls}}&  \multirow{2}{*}{current} & \multirow{2}{*}{20\%} & \multirow{2}{*}{20\%}\\
BR$(K_L\to \pi^0 ee)(-)$         & $(1.55^{+0.60}_{-0.48})\times 10^{-11}$    &  &  & \\
BR$(K_L\to \pi^0 \mu\mu)(+)$         & $(1.38^{+0.27}_{-0.25})\times 10^{-11}$    & \multirow{2}{*}{$ < 38\times 10^{-11}$ @$90\%$ CL} & \multirow{2}{*}{\cite{KTEV:2000ngj}} &  \multirow{2}{*}{current} & \multirow{2}{*}{20\%} & \multirow{2}{*}{20\%} \\
BR$(K_L\to \pi^0 \mu\mu)(-)$         & $(0.94^{+0.21}_{-0.20})\times 10^{-11}$    &  &  &  \\
\hline \hline
\end{tabular}}
\caption{\small
The SM predictions, current experimental values, and projected precisions. In the last three columns, ``current'' indicates that the measurement precision or the upper bound is kept to the current experimental value. 
\label{tab:data}}
\end{center}
\end{table}

\subsection{Comments on \texorpdfstring{$K_L \to e^+ e^-$}{KL->e+ee}} 

Following the discussion of the relative contributions of long versus short-distance contributions in $K_L\to  \mu\bar{\mu}$, here we would like to address the precision at which BR$(K_L\to e^+  e^-)$ has to be measured in order to have some interesting physics impact. We will repeat the $K_L\to \ell\bar{\ell}$ calculation  done in previous literature~\cite{DAmbrosio:1997eof, Valencia:1997xe, GomezDumm:1998gw, Isidori:1999ij, Isidori:2003ts, Hoferichter:2023wiy} (in particular Ref. \cite{Isidori:2003ts})  but focusing on the difference for $K_L\to e^+  e^- $, and  what it is interesting to measure.
Generically  for $K_L$ decaying into final lepton pairs we separate the long-distance two-photon intermediate state contribution, $A_{\gamma \gamma}$,  from a pure short distance one, $A_{\rm short}$:
\begin{equation}
A( K_L \to \ell  \bar\ell ) = \left( A_{\gamma \gamma} + {\rm Re} A_{\rm short} \right)
\bar \ell \gamma_5  \ell~.
\end{equation}
The long-distance two-photon intermediate state contribution, has a large imaginary piece, ${\rm Im} A_{\gamma\gamma}$.
It is then convenient to write $\Gamma(K_L \to \ell \bar\ell)$ in terms of the experimental value of 
$\Gamma(K_L \to \gamma\gamma)$ and  write the sum of the absorptive ($R_{\rm abs}$)  and dispersive ($ R_{\rm disp}$) contributions
\cite{Isidori:2003ts}
\begin{align}
\Gamma(K_L \to \ell \bar\ell) =&  \frac{2 \alpha_{\rm em}^2 r_l \beta_l  }{\pi^2} 
 \left[ R_{\rm abs} + R_{\rm disp} \right] \Gamma(K_L \to \gamma\gamma)~, \label{eq:dec1} \\
R_{\rm abs} =& ({\rm Im} C_{\gamma\gamma})^2 = \left[ \frac{\pi}{2 \beta_\ell }
\ln \frac{1-\beta_\ell}{1+\beta_\ell} \right]^2~, 
 \label{eq:dec2}  \\
R_{\rm disp} =&  \left[ \chi(\mu) -\frac{5}{2} +\frac{3}{2}\ln\left(\frac{m_\ell^2}{\mu^2}\right)
+ {\rm Re}  C_{\gamma\gamma} \right]^2~, 
  \label{eq:dec3} 
\end{align}
where $r_{\ell} = m_\ell^2/m_K^2$, $\beta_{\ell} = \sqrt{ 1-4r_\ell}$ and 
\begin{align}
 C_{\gamma\gamma} = \frac{1}{\beta_\ell} \left[ {\rm Li}_2\left( \frac{ \beta_\ell -1 }{ \beta_l+1} \right)
+\frac{\pi^2}{3} +\frac{1}{4} \ln^2 \left( \frac{ \beta_\ell -1 }{ \beta_\ell+1} \right) \right]~.
\end{align}
In eq. (\ref{eq:dec1}), compared to the muon mode,  $r_{e}$ suppresses BR$(K_L\to e^+  e^-)$ while enhances  $R_{\rm abs}$   and $ R_{\rm disp}$, as the small electron mass (with $\beta_\ell $ close to 1) makes large logarithms. Also, we need to have experimental information on 
the real coefficient  $\chi(\mu)=\chi_{\rm short}+\chi_{\gamma\gamma}(\mu)$ since then we can uncover 
short and long-distance information; $\chi_{\gamma\gamma}(\mu)$ depends on 
the $K_L \to \gamma\gamma$ form factor for which an analytic expression can be found in Ref.\cite{DAmbrosio:1997eof}. 
Based on the above arguments we have evaluated the experimental precision required for BR$(K_L\to e^+  e^-)$ in order to get a handle on  $\chi(\mu)$: Our finding indicates that percent level measurement (assuming the SM evaluation for $\chi_{\rm short}$) will achieve this goal.

\section{Global analyses}
\label{sec:global}
We present global analyses of the relevant rare kaon decay observables given in section~\ref{sec:2.1}, considering both the current experimental measurements and the projected situation as given in Table~\ref{tab:data}. For the projections, we do not assume an improvement in the precision of the theoretical predictions.

\subsection{Projections for NA62 and HIKE}
\label{sec:NA62_Hike}
\begin{figure}[tb]
\begin{center}
\includegraphics[width=0.48\textwidth]{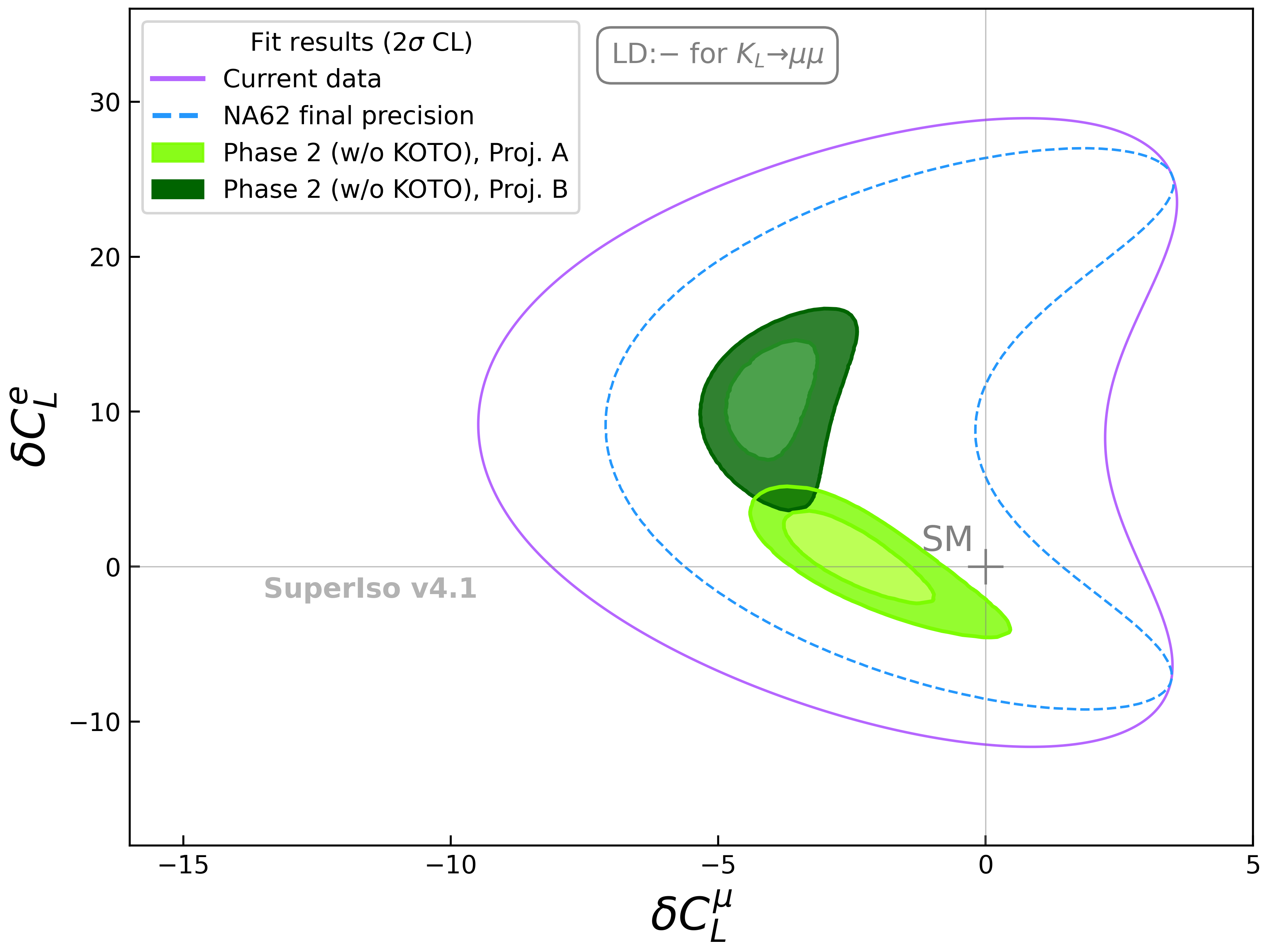}\quad
\includegraphics[width=0.48\textwidth]{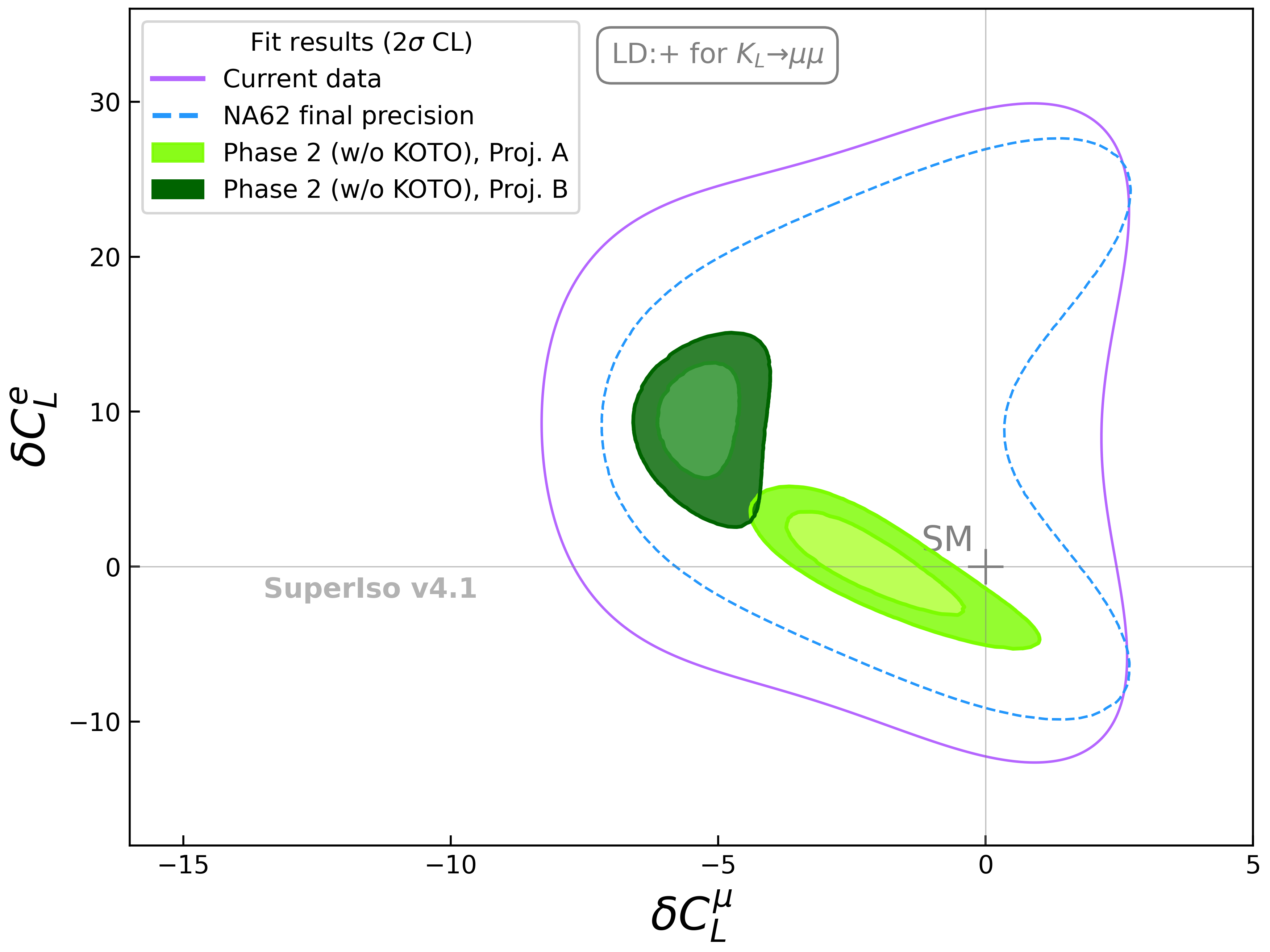}
\caption{\small Global fit to kaon observables with current data (purple solid contour) and final NA62 projection (dashed blue contour) at 95\% CL. 
The global fit for the projected HIKE sensitivity (corresponding to the penultimate column of Table~\ref{tab:data}) is shown at 68 and 95\% CL with  two shades of light (dark) green for scenario A (B). For further details, see the text. 
\label{fig:projections_WithoutKOTO}}
\end{center}
\end{figure}

The global fits to the rare kaon decays are given in the $\{\delta C_L^e, \delta C_L^\mu(=\delta C_L^\tau)\}$ plane. 
In Fig.~\ref{fig:projections_WithoutKOTO} the 95\% confidence level region of the fit to the current data is shown with the solid purple contour. The SM prediction of BR($K_L\to\mu\bar\mu$) depends on the long-distance (LD) contribution of $K_L\to\gamma \gamma$, and due to its sign ambiguity we consider fits both when the LD sign is negative (as given in the left plot of Fig.~\ref{fig:projections_WithoutKOTO}) as well as positive (the right plot of Fig.~\ref{fig:projections_WithoutKOTO}). 
In the fit to current data, the main constraining observable is the branching ratio of the decay $K^+\to\pi^+\nu\bar{\nu}$, but it does not distinguish between the electron and muon modes. It is the branching ratio of the decay $K_L\to\mu\bar{\mu}$ that breaks this degeneracy. Overall, these two decays are currently the main constraining observables in the global fit (a detailed investigation of the impact of individual observables on this fit can be found in Ref.~\cite{DAmbrosio:2022kvb}).

For the projections, an obvious benchmark point is the final precision of NA62, at which point BR($K^+ \to \pi^+ \nu\bar\nu$) will be measured with 20\% uncertainty~\cite{KM:pricom}. For this case, we assume that the future central value remains the same as the current measurement. The 95\% confidence level contour of the fit is shown in Fig.~\ref{fig:projections_WithoutKOTO} with the dashed blue outline. The shrinking of the $2\sigma$ region compared to the current fit is only due to the better experimental precision of BR($K^+ \to \pi^+ \nu\bar\nu$), where the rest of the data are kept unchanged.

We also present the global fit results for projections based on HIKE Phase 2 target precision (penultimate column in Table~\ref{tab:data}). In this case, the precision of BR($K^+\to \pi^+ \nu\bar\nu$) will be improved to 5\%, and the precision of BR($K^+\to\pi^+ \ell\bar\ell$) is expected to be improved by more than a factor of two. Most significantly, the challenging ultra-rare decays $K_L\to\pi^0 \ell\ell$ will be measured with 20\% precision. 
Here, we assume there will be no improvement in the measurement of BR($K_L\to\pi^0\nu\bar\nu$), although within the same time frame, the KOTO-II collaboration is expected to have observed and measured this decay (see Section~\ref{sec:koto} below for fit projections including KOTO-II precision).

\begin{figure}[t!]
\begin{center}
\includegraphics[width=0.48\textwidth]{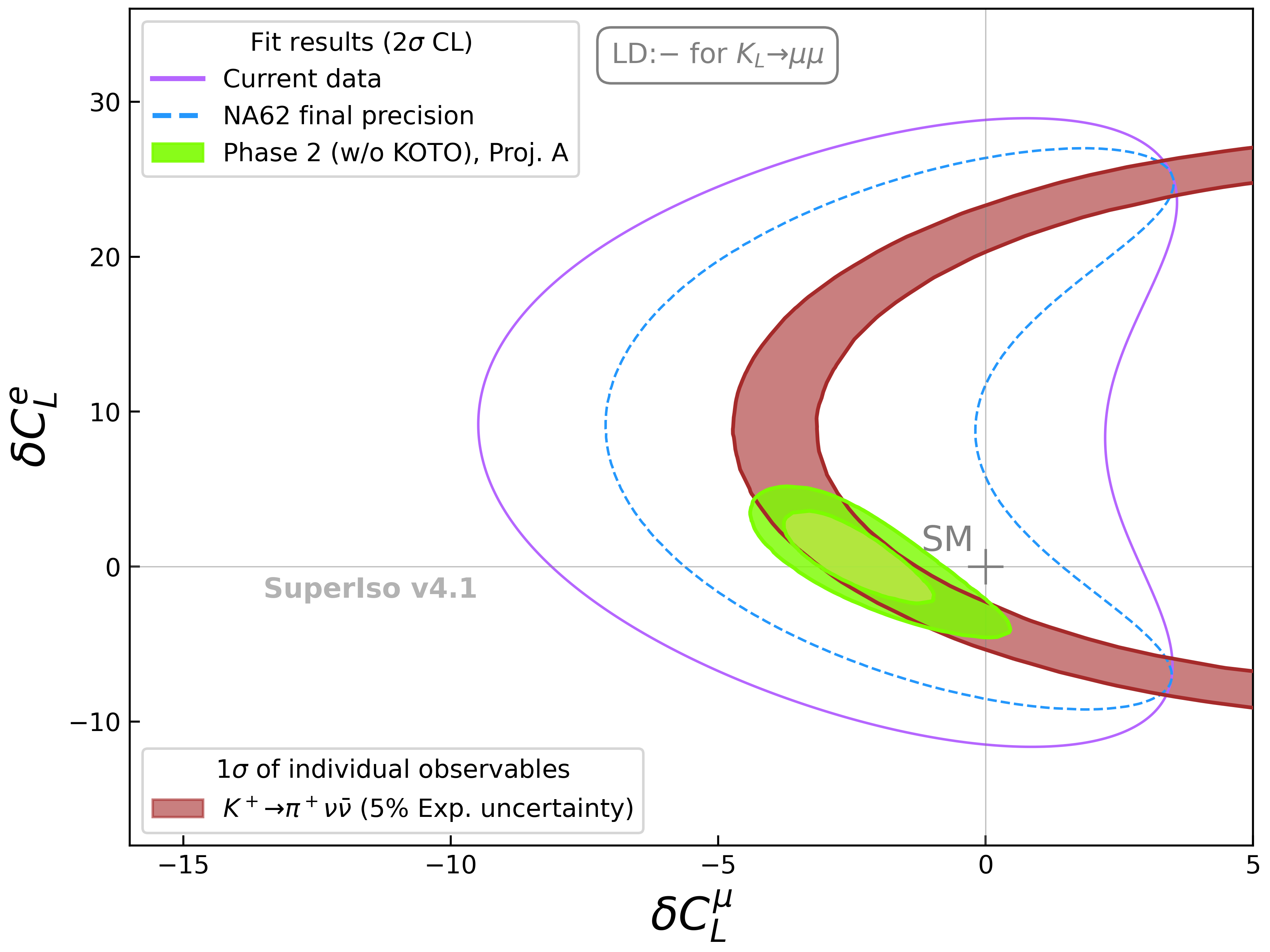}\quad
\includegraphics[width=0.48\textwidth]{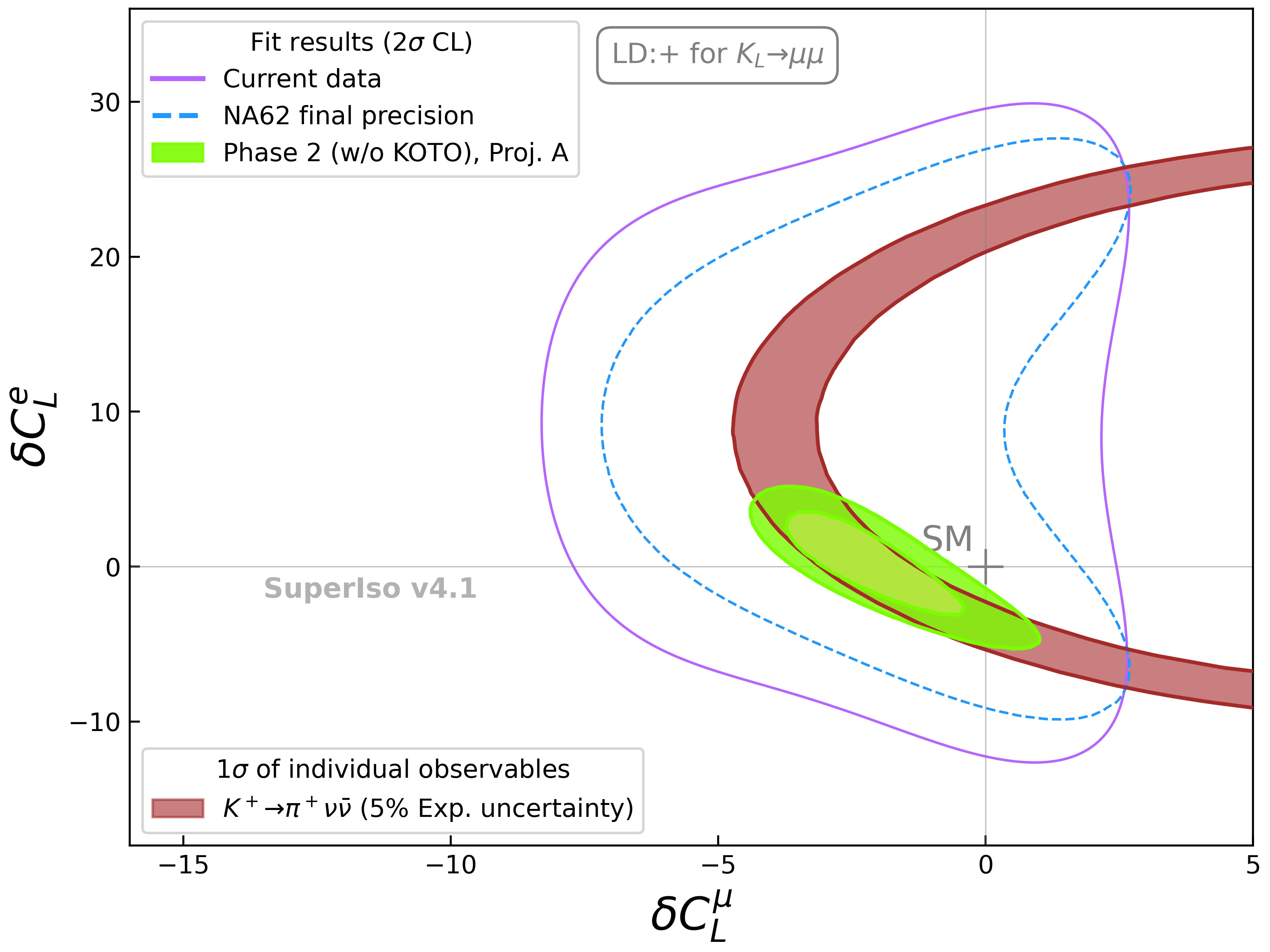}
\includegraphics[width=0.48\textwidth]{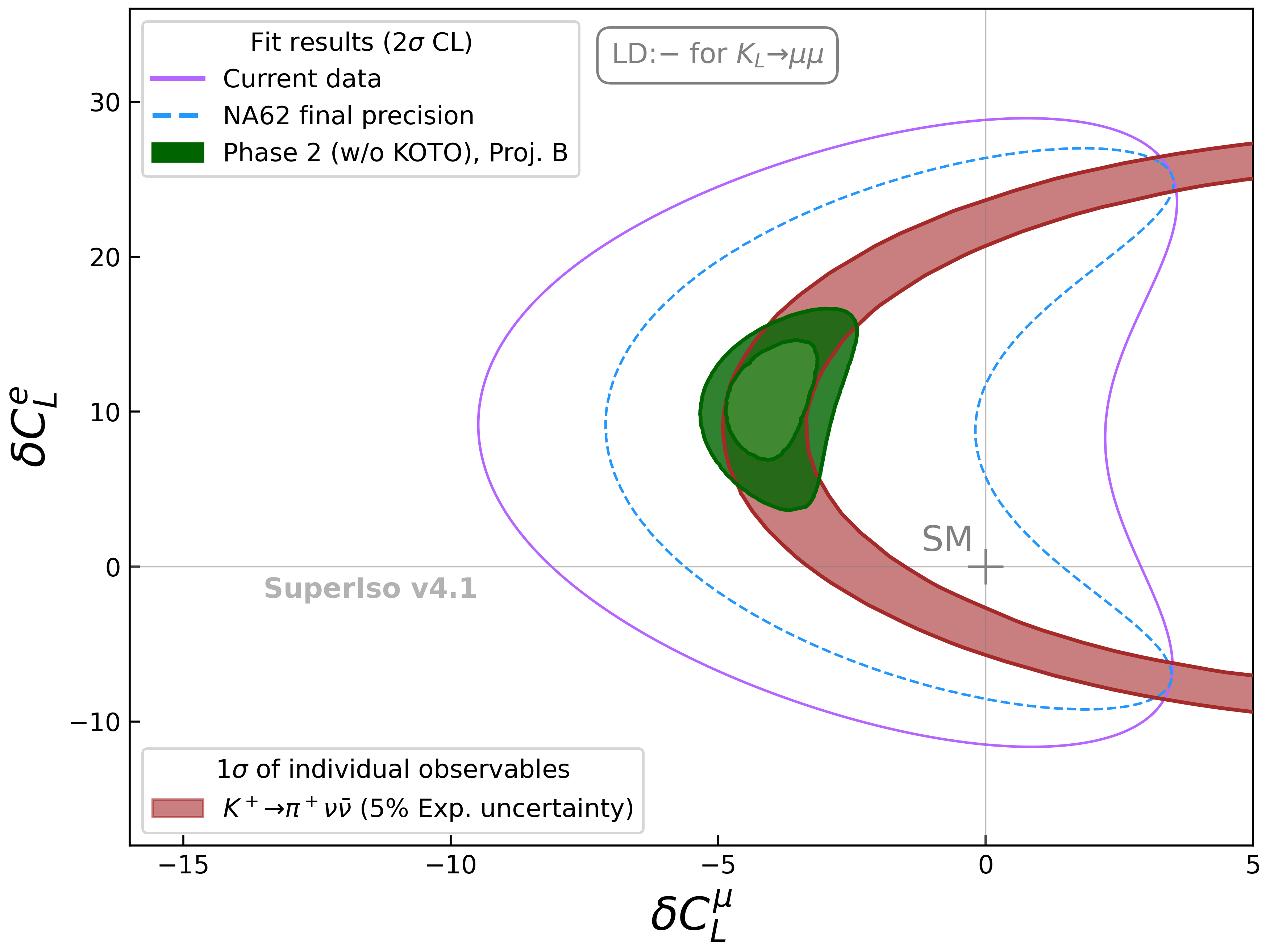}\quad
\includegraphics[width=0.48\textwidth]{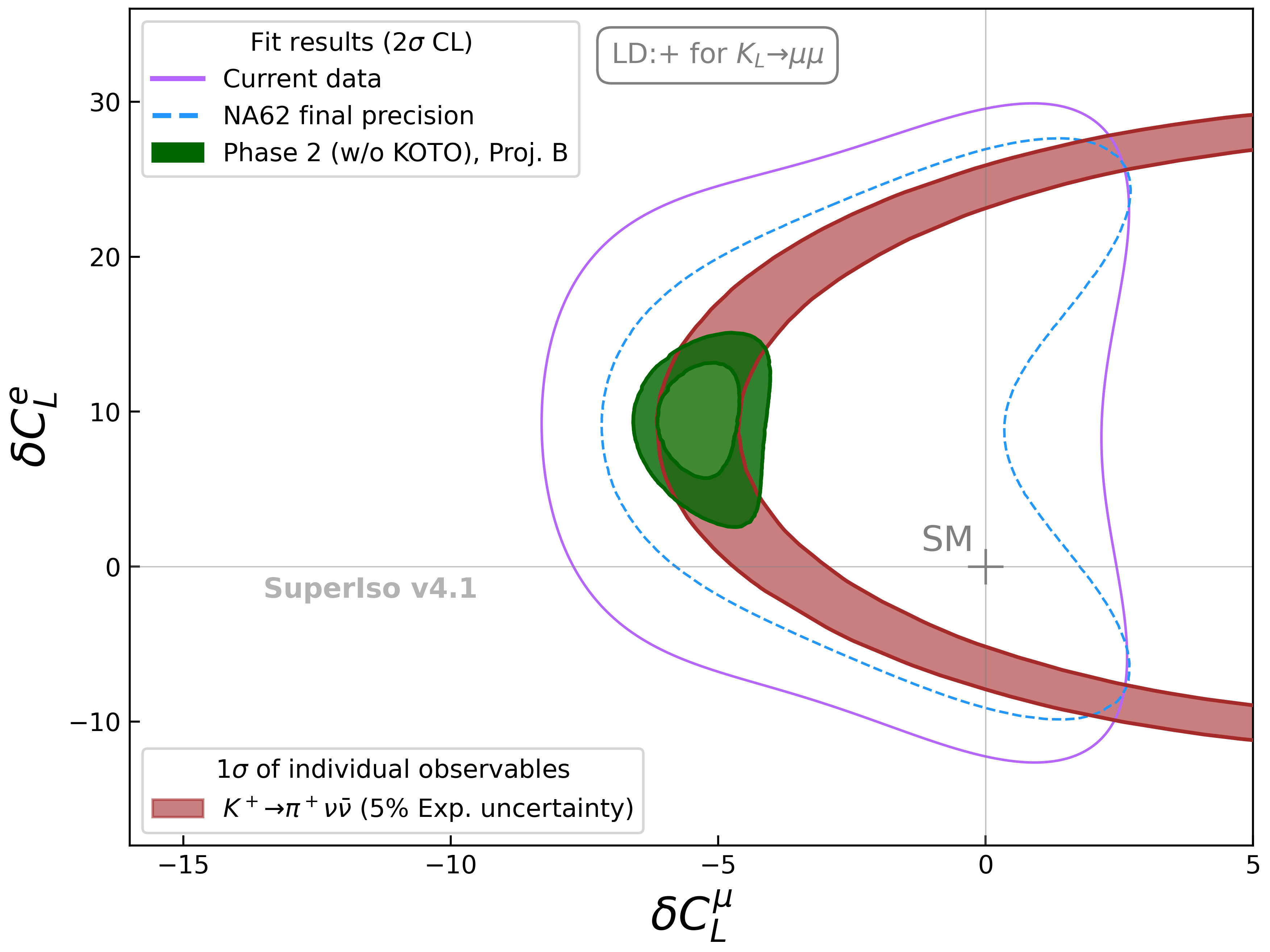}
\caption{\small Impact of $K^+\to\pi^+\nu\bar\nu$ on the projected fits for HIKE sensitivity. See the text and the caption of Fig.~\ref{fig:projections_WithoutKOTO} for further detail.
\label{fig:detalied_WithoutKOTO}}
\end{center}
\end{figure}

The target precision of the projected data can be read off from Table~\ref{tab:data}, however, since currently some observables have not been measured, for the central values we assume projections for two different scenarios:
\begin{itemize}
\item \textbf{Projection A:} For the observables that have been measured we consider their current central values and for the others we consider their SM predictions.

\item \textbf{Projection B:} We project all central values based on the best-fit values of the Wilson coefficients from the fit to the current data.
\end{itemize}
The results of the fits at 68 and 95\% confidence levels for future HIKE sensitivity are given in Fig.~\ref{fig:projections_WithoutKOTO} for Projection~A with the two shades of light green and for Projection~B with the two shades of dark green.
For Projection~A, we find consistency with the SM at the $3\sigma$ level. This is expected since the decays that have not yet been experimentally observed are assumed to be measured with central values corresponding to their SM predictions.  
For Projection~B, there  is a clear departure from the SM, which reflects the fact that the current best-fit point (with which the yet unobserved decays have been projected) does not coincide with the SM. There is only a small overlap at $2\sigma$ level between the two scenarios for the case of LD$-$ contributions to $K_L\to\mu\bar\mu$, this overlap is even smaller for the case of LD$+$.   

\begin{figure}[t!]
\begin{center}
\includegraphics[width=0.48\textwidth]{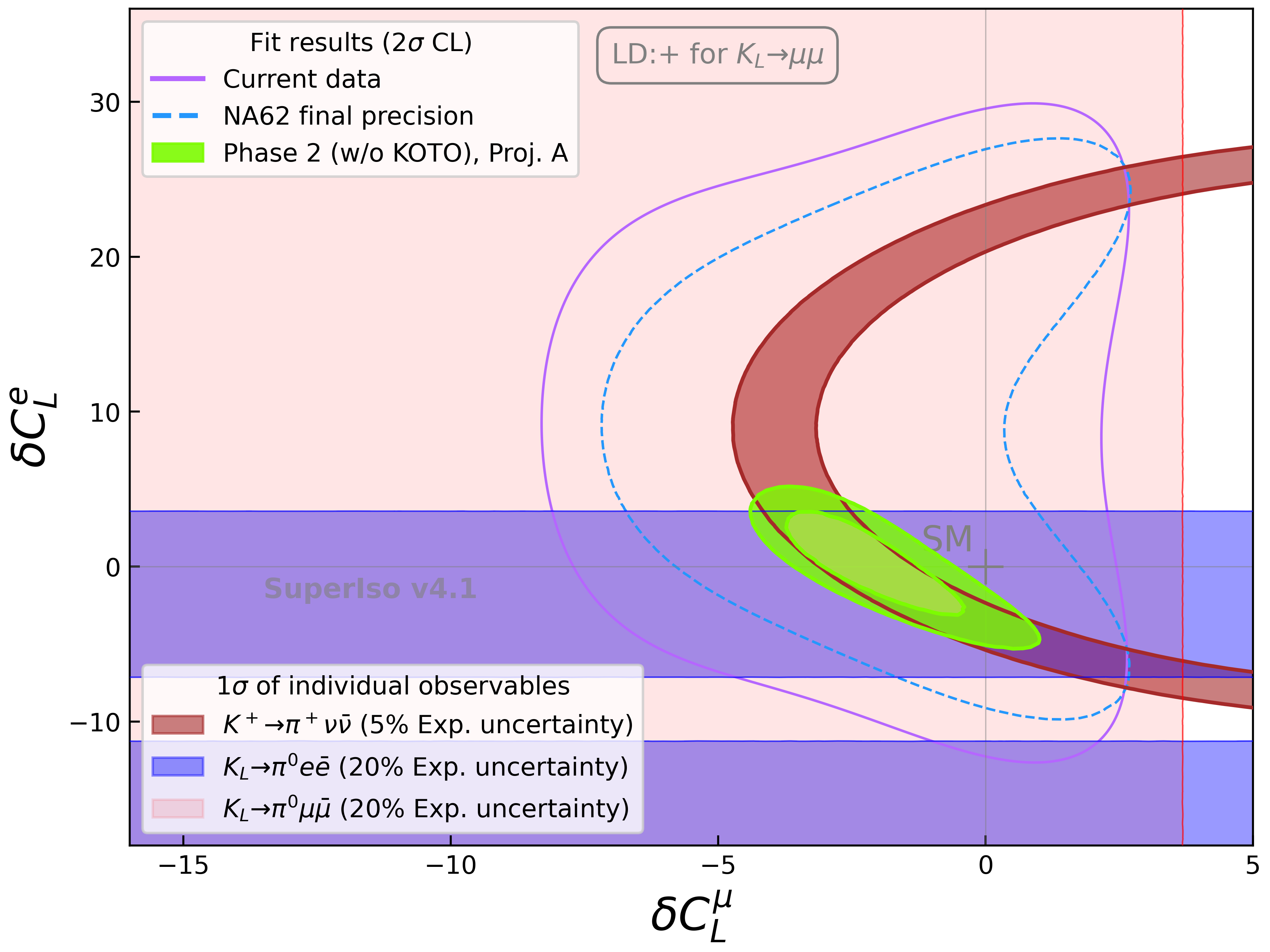}\quad
\includegraphics[width=0.48\textwidth]{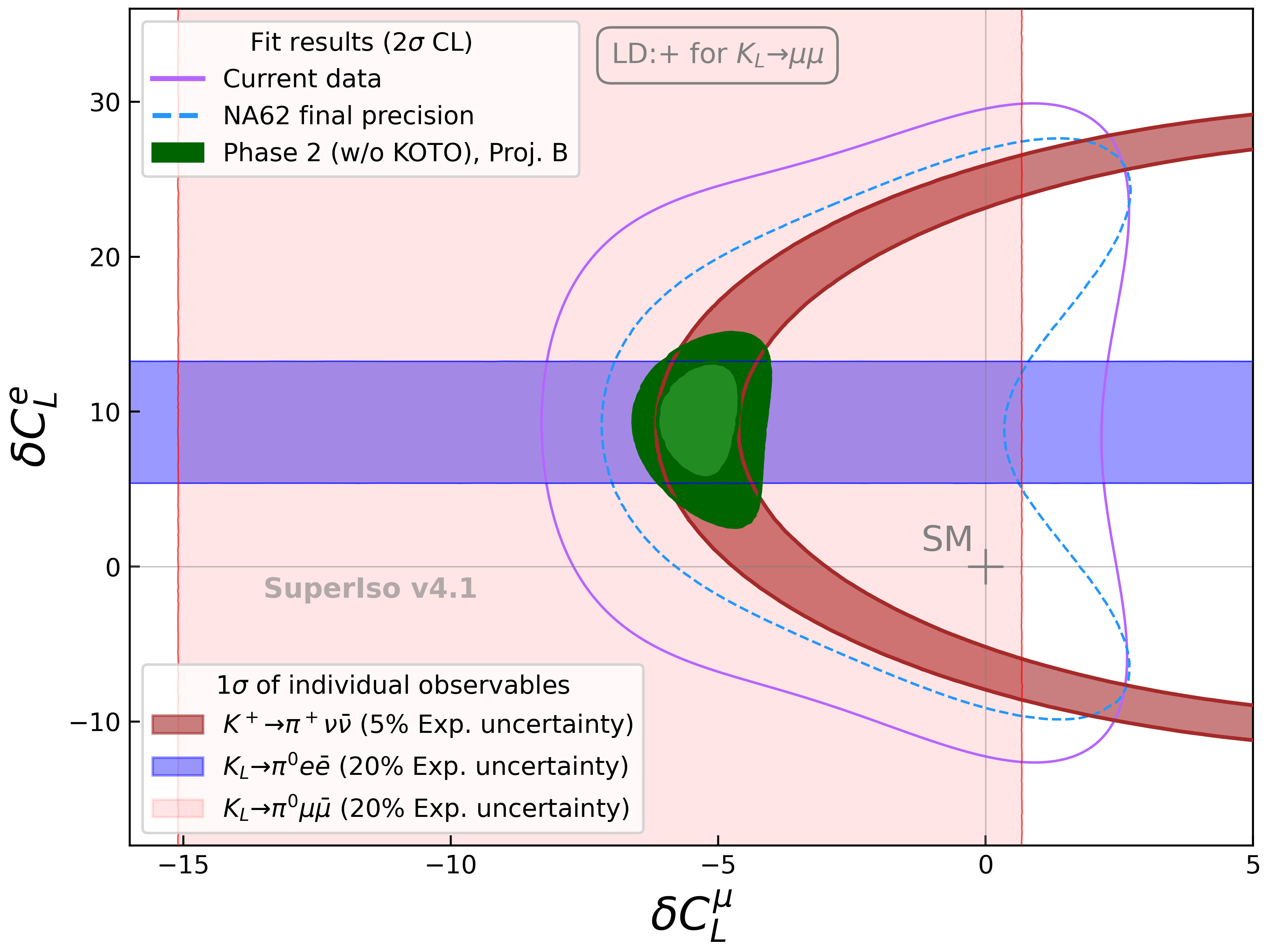}
\caption{\small The impact of $K_L\to \pi^0 e^+e^-$ and $K^+ \to \pi^+ \nu \bar\nu$ on the projected fits for HIKE sensitivity. The left (right) plot corresponds to the upper (lower) right plot of Fig.~\ref{fig:detalied_WithoutKOTO}.
\label{fig:separated_WithoutKOTO}}
\end{center}
\end{figure}

In a similar way to the fit to current data, also for the projections one of the main constraining observables is BR($K^+\to\pi^+\nu\bar\nu$) which is projected with 5\% uncertainty. This is visible in Fig.~\ref{fig:detalied_WithoutKOTO}, where we show the two different projections in separate plots (for either sign of LD contributions to $K_L\to\mu\bar\mu$). The brown half-oval shape which indicates the $1\sigma$ bound due to BR($K^+\to\pi^+\nu\bar\nu$) coincides nicely with the $1\sigma$ CL of the fit. However, clearly there are other observables that have a strong impact on the fits. While in the fit to current data, the main observable disentangling $C_L^e$ and $C_L^\mu$ is in the muon mode via BR($K_L\to \mu\bar\mu$), for the projections this is taking place in the electron mode via BR($K_L\to\pi^0e\bar e$). In Fig.~\ref{fig:separated_WithoutKOTO}, in addition to the bound due to BR($K^+\to\pi^+\nu\bar\nu$) the impact of BR($K_L\to\pi^0 \ell\bar\ell$) is also shown at $1\sigma$ level (for the case of LD$+$ contributions to $K_L\to\mu\bar\mu$). While for both projections the strong impact of $K_L\to \pi^0 e\bar e$ is visible, for Projection~B (right plot) it is more prominent; this is by construction,  as all the projected data have been produced with the  same (current) best-fit point. 
For both projections, the main two constraining decays are $K^+\to\pi^+\nu\bar\nu$ and $K_L\to\pi^0 e\bar e$. 

\subsection{Impact of KOTO-II}
\label{sec:koto}
\begin{figure}[t!]
\begin{center}
\includegraphics[width=0.48\textwidth]{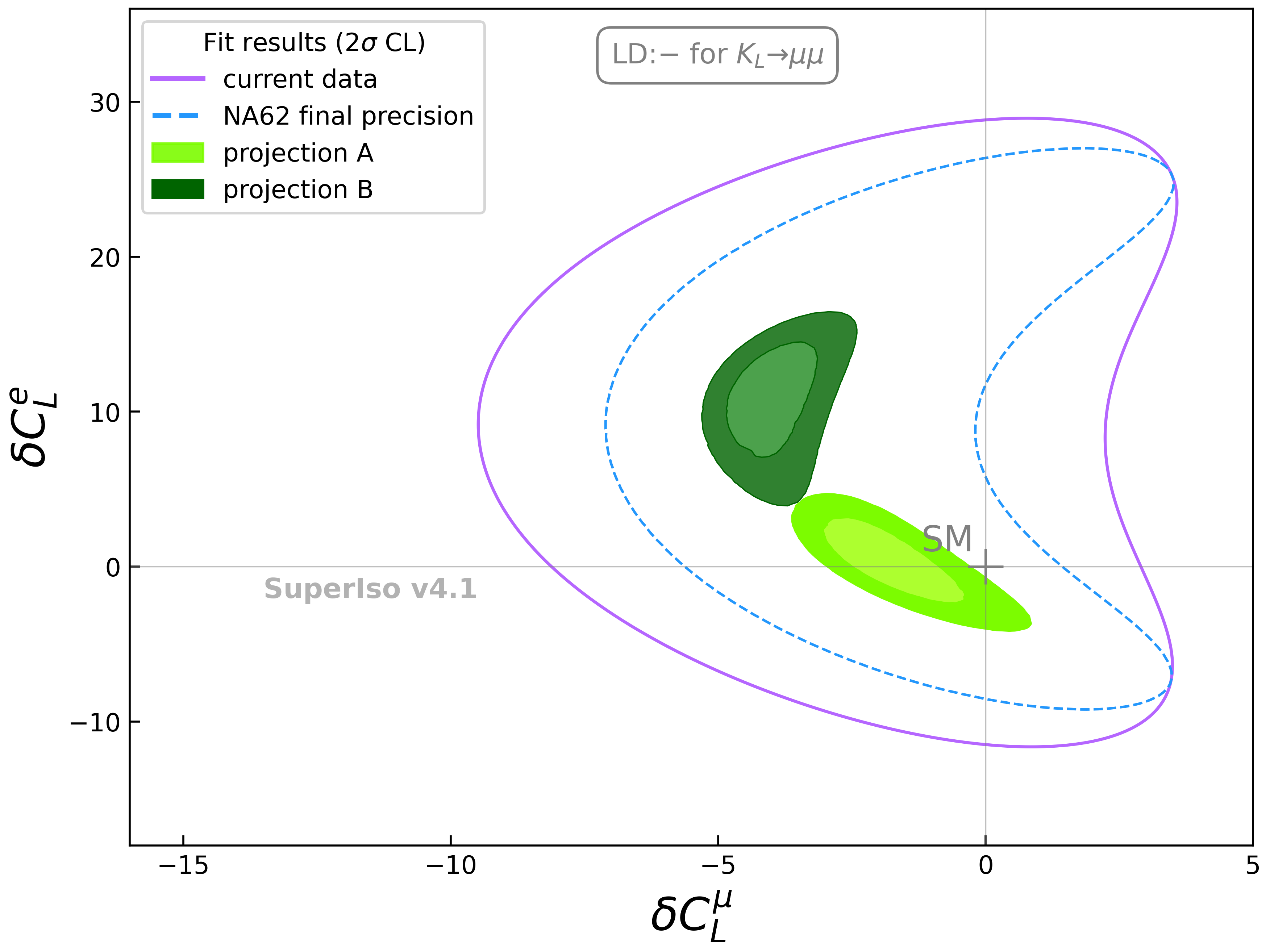}\quad
\includegraphics[width=0.48\textwidth]{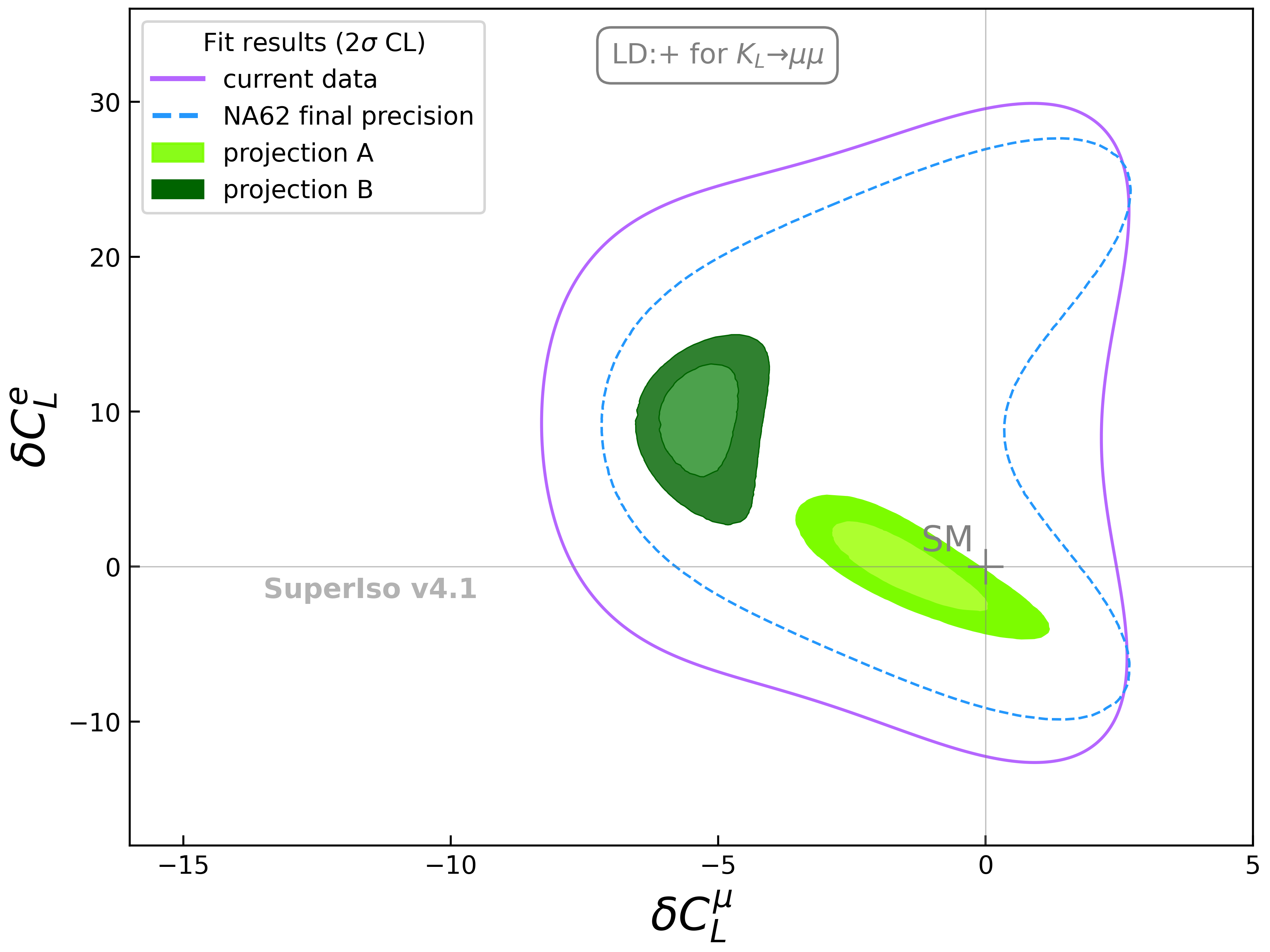}
\caption{\small 
Projected fits for HIKE sensitivity together with KOTO-II (as given in the last column of Table~\ref{tab:data}).
\label{fig:projections_WithKOTO}}
\end{center}
\end{figure}

\begin{figure}[H]
\begin{center}
\includegraphics[width=0.48\textwidth]{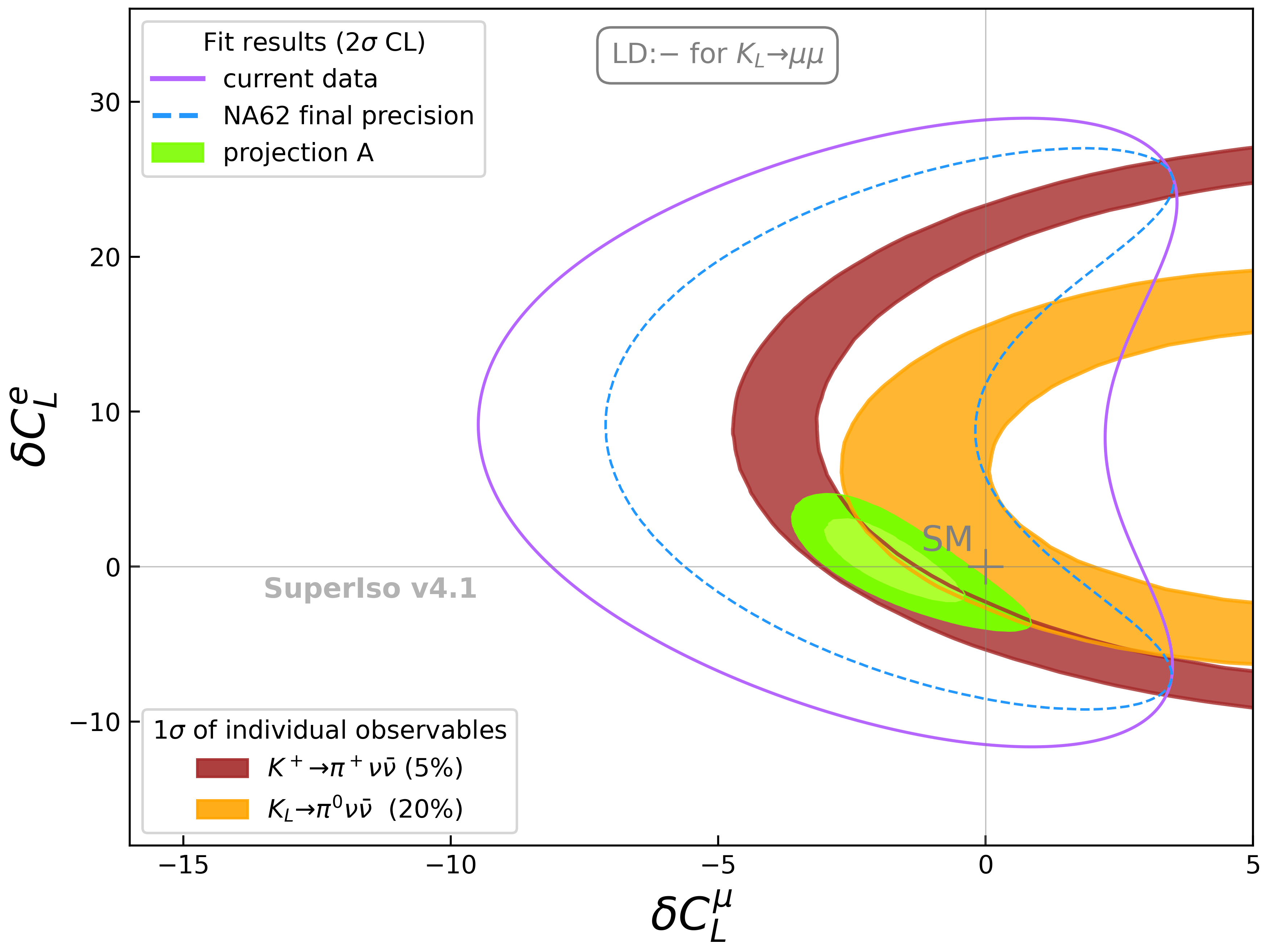}\quad
\includegraphics[width=0.48\textwidth]{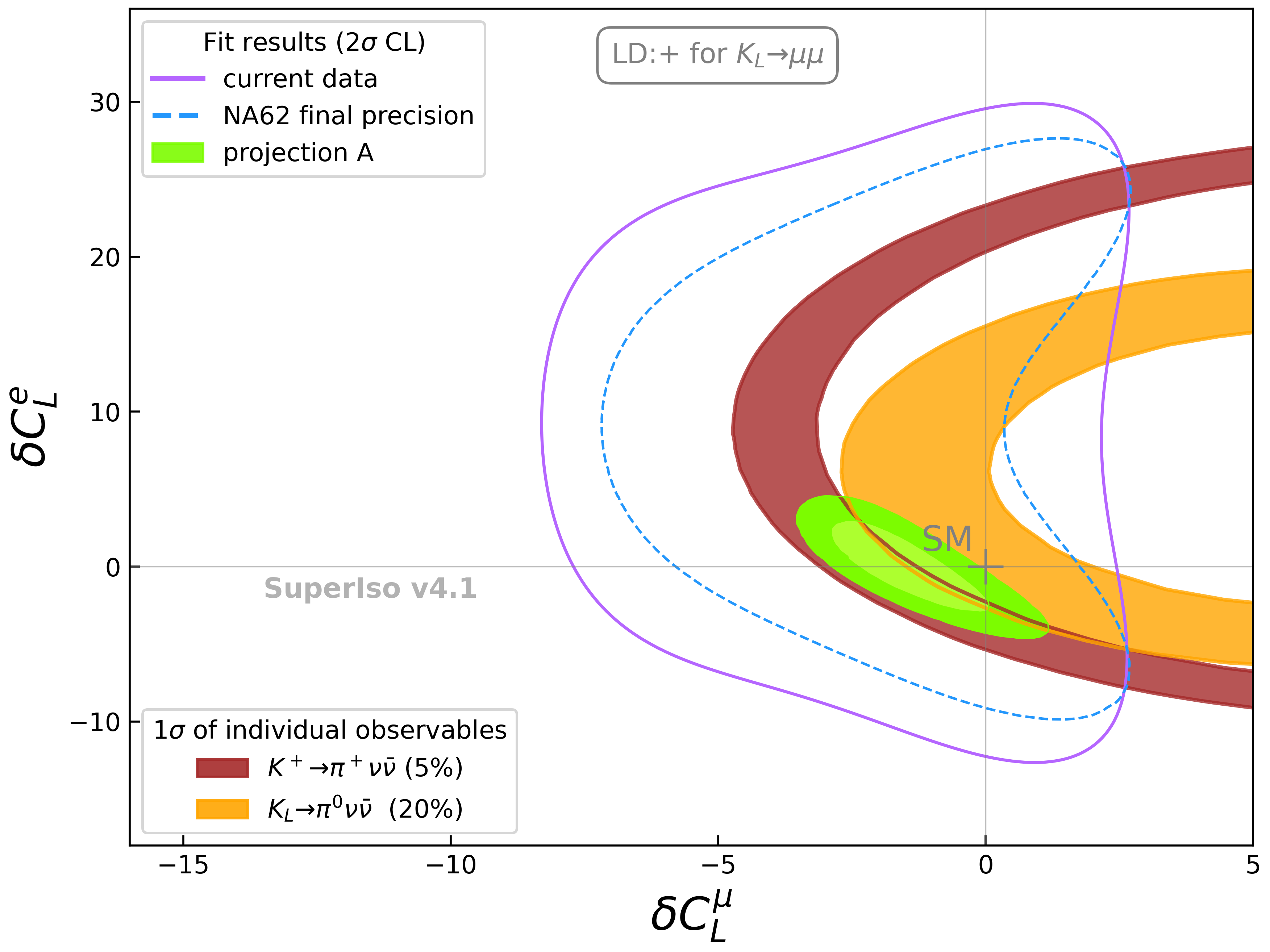}
\includegraphics[width=0.48\textwidth]{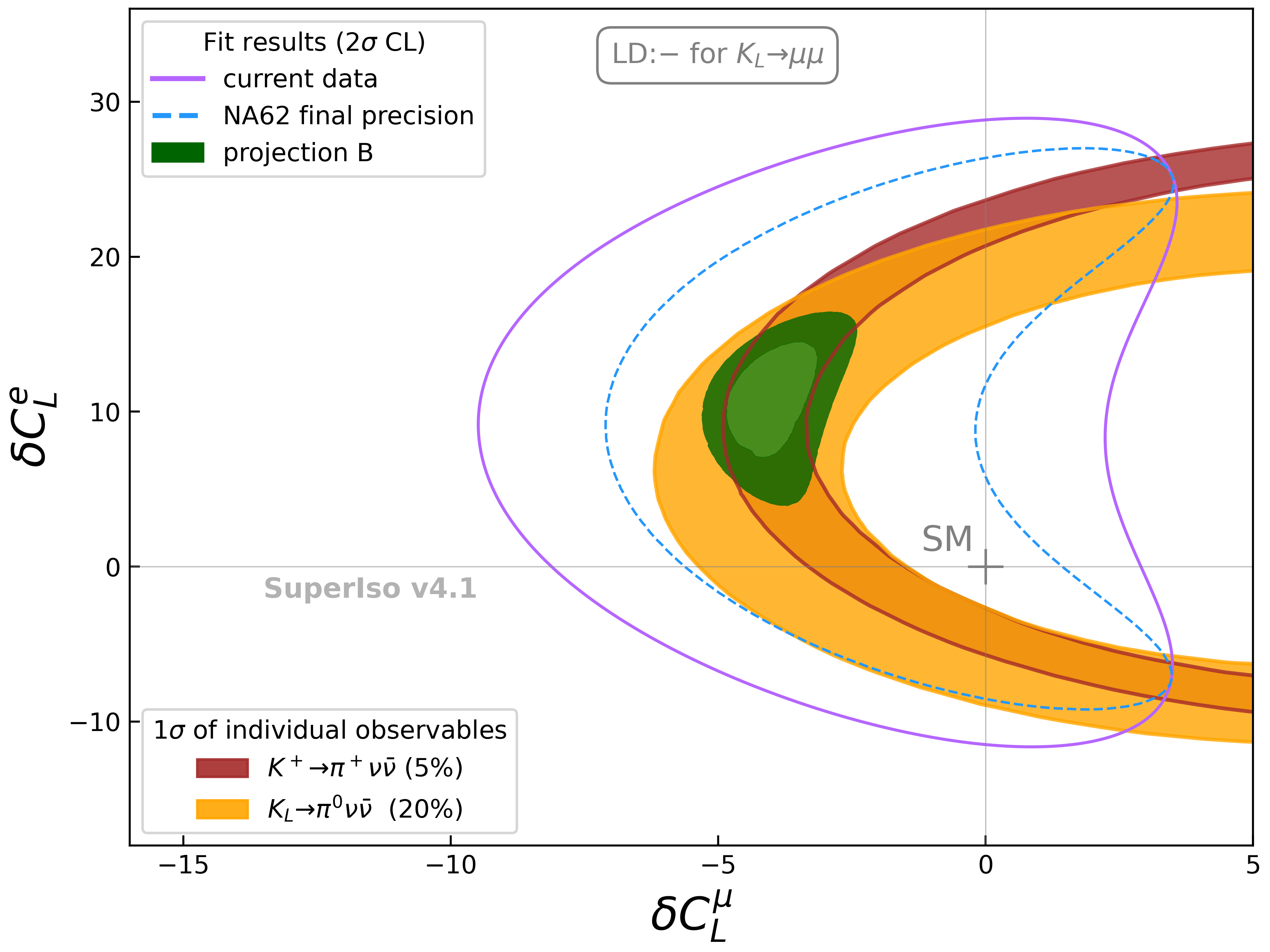}\quad
\includegraphics[width=0.48\textwidth]{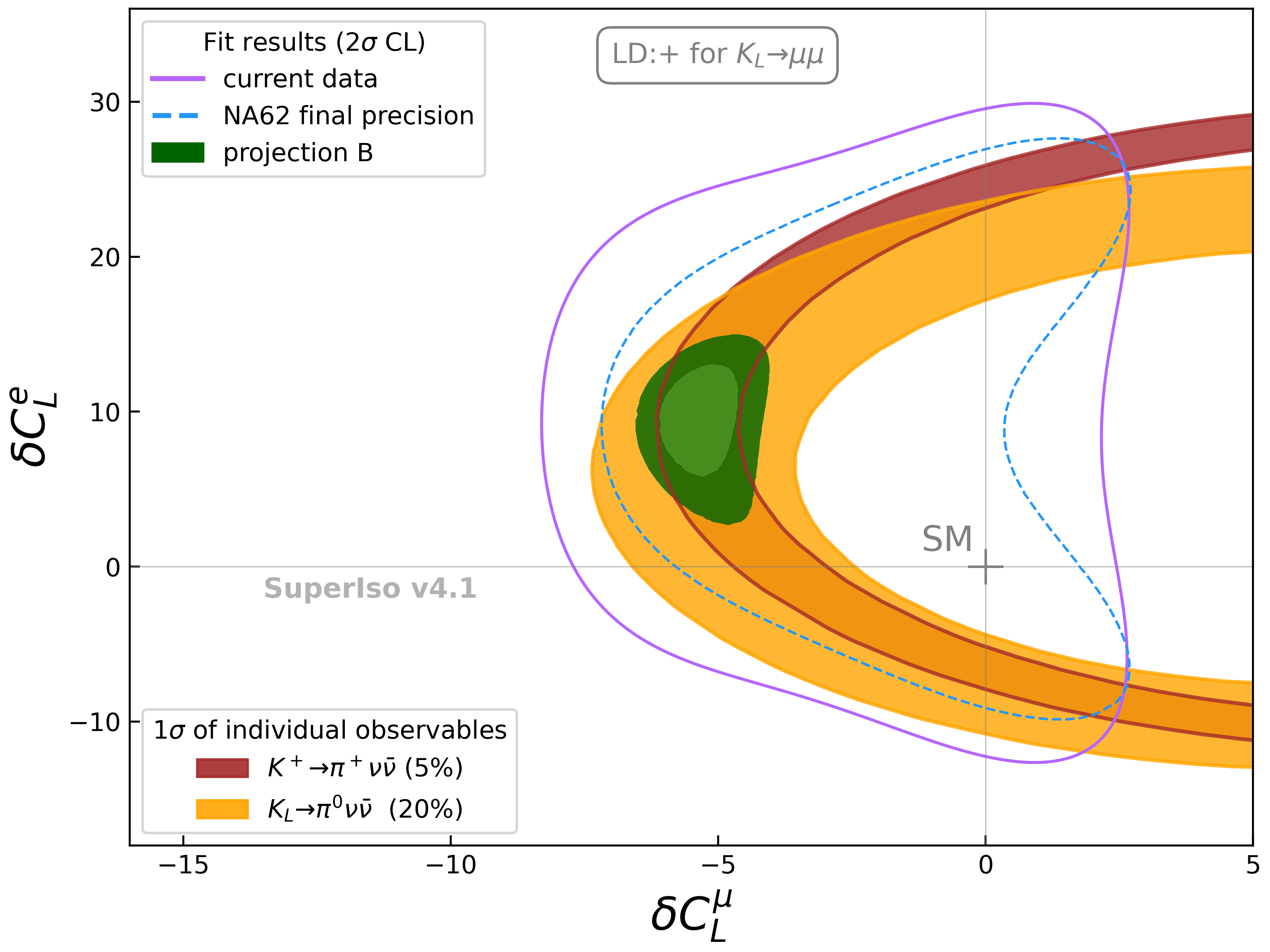}
\caption{\small Impact of $K^+\to\pi^+\nu\bar\nu$ and $K_L\to\pi^0\nu\bar\nu$ on the projected fit for HIKE and KOTO-II sensitivities. 
\label{fig:detailed_WithKOTO}}
\end{center}
\end{figure}

In this subsection, in addition to the future HIKE sensitivity, we also consider 20\% precision measurement of $K_L\to\pi^0\nu\bar \nu$  by KOTO-II. 
We use the same setup as described in the previous subsection for only HIKE sensitivity.
The projected fits are displayed in Fig.~\ref{fig:projections_WithKOTO} with the light and dark green shades for Projections A and B, respectively. The two scenarios no longer have any overlap at the $2\sigma$ level, and in the case of LD$+$ contributions to $K_L\to\mu\bar\mu$, the distinction of the two scenarios is more pronounced. 
Comparing Fig.~\ref{fig:projections_WithKOTO} with Fig.~\ref{fig:projections_WithoutKOTO}, the fits now indicate a more constrained region which is due to the additional bound from KOTO-II. Furthermore, Projection A is slightly more consistent with the Standard Model because in this scenario it is assumed that the central value of BR($K_L\to\pi^0\nu\bar \nu$) will match the SM prediction.  The impact of the 20\% KOTO-II measurement is demonstrated in Fig.~\ref{fig:detailed_WithKOTO} with the yellow half-oval shape where the $1\sigma$ bound of BR($K_L\to\pi^0\nu\bar \nu$) is shown.

\section{Conclusions}
\label{sec:conc}

In this work, we investigated the potential physics reach of rare kaon decay measurements with the future HIKE programme at CERN. We studied the constraining power of rare kaon decays on short-distance physics in a model-independent approach for beyond the Standard Model scenarios with lepton flavour universality violation. 

We displayed the fit to current data and also made  projections for the final precision of NA62 experiment considering 20\% uncertainty for the golden channel $K^+\to\pi^+\nu\bar\nu$. 
Furthermore, we performed projected fits for HIKE Phase 2 target precision. In addition to the improvements in the golden channel and $K^+\to\pi^+\ell\bar\ell$ that have been measured, HIKE will provide the first measurements of channels that have not yet been observed.


\noindent Therefore, we considered two different scenarios: A) considering future measurements for yet unobserved decays to match the Standard Model predictions, B) all observables are projected to match the best fit with current data.  
We showed that future measurements of rare kaon decays by HIKE offer very powerful constraints on lepton flavor universality violating new physics effects, leading to clearly different results.

Furthermore, we highlighted the decays that are most impactful in the global fit and demonstrated that besides the golden channel, the $K_L\to\pi^0\ell\bar\ell$ decays, especially in the electron channel put strong constraints on new physics scenarios. This is in contrast to the fits with current data where the strongest constrain besides $K^+\to \pi^+\nu\bar\nu$ is from $K_L\to \mu\bar\mu$ (which disentangles the electron and muon contributions).

Moreover, we performed global fits including the future $K_L\to\pi^0 \nu\bar\nu$ measurement with 20\% precision by KOTO-II in addition to future HIKE measurements resulting in more constrained fits. However, from the projected fits it is clear that HIKE Phase 2 offers strong sensitivity to new physics, even without improvements in the experimental measurement of $K_L\to \pi^0\nu\bar\nu$. 

To clearly highlight the impact of future experimental measurements, we have not taken into account here any improvement in the theoretical precision of the considered observables. Our results are therefore rather conservative as it is reasonable to expect better theoretical precisions in the calculation of the relevant observables, resulting in even stronger constraints and bounds for new physics.

\section*{Acknowledgements}
We would like to thank Cristina Lazzeroni, Gino Isidori, Marc Knecht, Urs Wiedemann and Karim Massri for fruitful discussions and suggestions. 

\bibliographystyle{JHEP}
\bibliography{kaon}

\end{document}